# Development of BPM electronics for PIP-II at Fermilab*

Shengli Liu[†], Nathan Eddy, A. Semenov, Brian Fellenz
Fermilab, Batavia, USA

*Abstract*

This paper presents the uTCA4.0-based BPM electronics for PIP-II, featuring four 250 MSPS ADCs and a Xilinx UltraScale+ MPSoC FPGA with 10 GbE uplink. Design elements include signal conditioning, clock, and thermal management. The FPGA performs signal processing, time tagging, digital down-conversion, and phase drift compensation. Position and phase resolution, and thermal stability — is validated through dedicated testing.

## INTRODUCTION

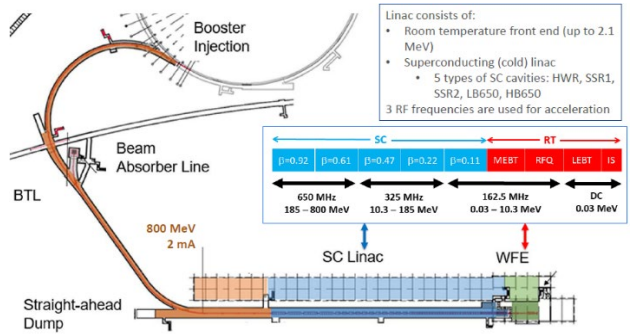

Figure 1: Beamline layout of the PIP-II SC Linac and BTL

PIP-II (Proton Improvement Plan-II) is Fermilab's next-generation high-intensity superconducting proton LINAC, replacing the aging 400 MeV linac with an 800 MeV H⁻ beam injected into the Booster [1]. As shown in Fig. 1, the beamline includes the warm front-end (WFE)—comprising the ion source, LEBT, RFQ, and MEBT—and operates at three RF frequencies: 162.5 MHz (RFQ and early SC sections), 325 MHz, and 650 MHz. The H⁻ beam is bunched at 162.5 MHz, chopped into pulses up to 0.55 ms (with future CW capability), and operates at up to 20 Hz repetition.

PIP-II will deploy 126 BPMs across the WFE, SC LINAC, and BTL. Warm and cold BPMs measure horizontal/vertical position, beam phase relative to the RF, and relative intensity, with performance requirements of: 10 μm position resolution, 0.1 mm accuracy, 1% intensity resolution, 0.3° phase resolution, and 1° phase stability. The 56 BTL BPMs measure a single transverse coordinate, totalling 392 BPM signals for processing.

## BPM Data Acquisition System

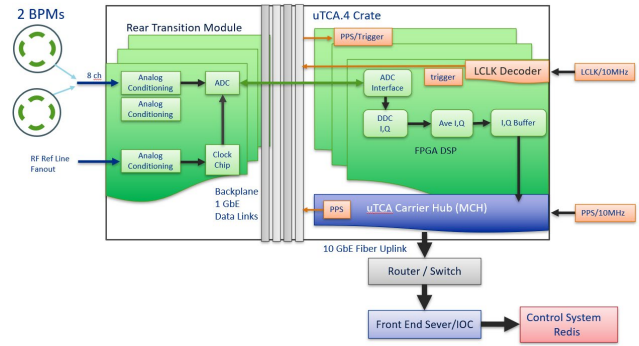

Figure 2: PIP-II's uTCA BPM System Architecture

The latest uTCA.4 platform offers cutting-edge technologies, including high-speed Ethernet for data transfer, hot-swappable modules for high availability, and IPMI-based platform management. The PIP-II BPM data acquisition system is built on this platform and consists of a uTCA Carrier Hub (MCH), up to 12 Advanced Mezzanine Cards (AMCs), and 12 Rear Transition Modules (RTMs), as shown in Fig. 2.

Each AMC/RTM pair accepts 8 analog inputs, enabling a 12-slot chassis to process up to 96 signals. The analog and digital sections are connected via the zone-3 connector, eliminating the need for extensive front-panel RF cabling typical in VME systems. Each RTM receives a 162.5 MHz RF reference signal distributed via a low-drift RF line in the beam tunnel.

A 10 MHz clock and an encoded clock (LCLK) are input through one AMC front panel and distributed over MLVDS buses (ports 17–20) on the backplane. LCLK is decoded on each AMC to extract trigger and other event signals. The White Rabbit LEN module supplies a 10 MHz and PPS signal via the MCH which distributes these clocks through TCLKA and TCLKB lines in a star topology. Together this provides single 162.5MHz bucket resolution.

Each AMC sends processed BPM data through backplane port 0 via 1 GbE to the MCH, which aggregates and forwards via 10 Gbps uplink to a front-end (FE) server—a 1U industrial Linux computer. The FE server further processes and publishes BPM data to PIP-II's control system using the REDIS distributed database.

### RTM: Analog Signal, Clock, and ADC

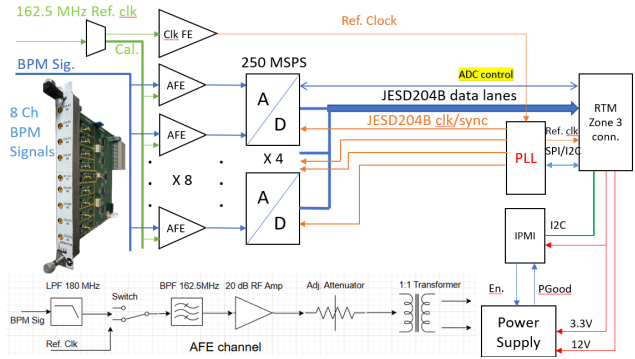

Fig. 3: PIP-II BPM RTM board

As seen the fig. 3, The RTM's Analog Front-End (AFE) channel includes a high-isolation RF switch, a SAW bandpass filter, and a fixed 20 dB RF amplifier. The clock channel uses the same AFE as the BPM signal path to ensure shared thermal drift characteristics, preserving phase stability. However, the drift of the PLL needs an active self-calibration to be performed.

The RTM integrates TI's 250 MSPS ADS42JB69 ADC with a JESD204B interface, necessitating the use of TI's JESD204B-compliant LMK04832 clock chip to meet deterministic latency requirements. Signal integrity for JESD204B was a key focus in PCB layout. Since the ADC-to-FPGA distance was unknown during design, a worst-case loss scenario was assumed, and the Megtron-6 material was chosen over standard FR4 to minimize signal degradation. With a power requirement of approximately 20 W, a dedicated heat sink was implemented for thermal management.

### AMC: FPGA, Digital Signal Processing

The Vadatech AMC566 [2] was selected to meet PIP-II's high-speed data acquisition requirements. Based on the Xilinx UltraScale+ XCZU7EV MPSoC FPGA, the AMC566 features four ARM cores running embedded Linux, abundant programmable logic resources, dual GbE, 16 GB DDR4 memory, and flexible boot modes. It supports the JESD204B interface, enabling seamless integration with the custom-designed RTM.

Key programmable logic (PL) modules include the ADC interface, digital down-conversion (DDC), and a data packing and time-tagging unit. The ADC and DDC leverage Xilinx JESD204B, NCO, and CIC IP cores. The data packing module outputs two AXI streams: high-speed raw ADC data (buffered in PL-side DDR4) and decimated I/Q data (transferred directly to the processing system). Either stream can be routed to the PS via a 32-bit, 100 MHz DMA channel (3.2 Gbps bandwidth).

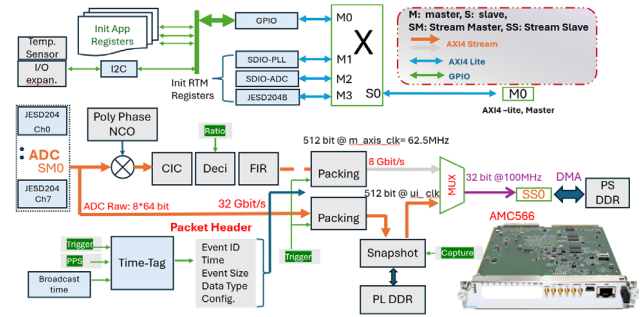

Fig. 4: FPGA: digital signal processing

Data is packetized on beam trigger events. At a maximum trigger rate of 20 Hz and 1% beam duty cycle, the throughput is 320 Mbps for raw ADC data and 20 Mbps for decimated I/Q data (assuming a decimation factor of 16). Each packet includes a 512-bit header with metadata such as device ID, event ID, time tag, packet size and attributes, and diagnostic info, structured for compatibility with a shared PS software stack.

Sub-nanosecond time-tagging will be enabled via White Rabbit hardware.

### Performance Benchmarks

*Test Setup*

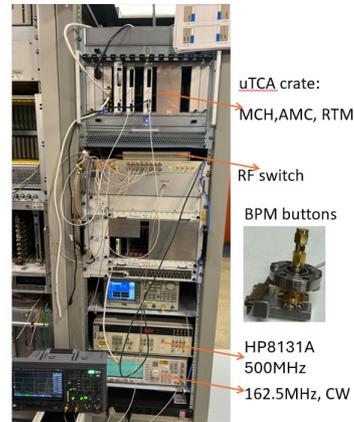

Fig. 5: Test setup: uTCA chassis, RF source, Fast pulser, BPM button

BPM buttons are used to create bunch signals. See Fig. 5. A fast pulser HP8131A-500MHz is triggered by the 162.5MHz sinewave to excite the button by feeding a 5V, 700 ps FWHM signal to a microstrip transmission line running through the surface of the BPM button. So the bunch signal is synchronized to the 162.5MHz RF source to simulate the real beam signal. The result button signal is about ~ 120 mV (-14.4dBm) in amplitude, 700 ps in

FWHM. Additionally, to the BPM button signal, the 162.5 MHz sine wave is also used for characterizing the electronics.

*Input range, Linearity and S/N:*

The RTM Rev5 prototype is measured to be 16 dB gain of AFE channel. ADC reaches the full range at -4.4 dBm sinewave input with good linearity due to a high linear Mini-circuit amplifier ADL5536 with the expense of higher power consumption. To decide a useful input range, the position errors need to be considered. The useful dynamic input range is measured to be -45 dBm to -4.4 dBm with X, Y error, and resolution less than 10 um. The nominal PIP-II BPM signal strength is above -10 dBm and is stable at each BPM location.

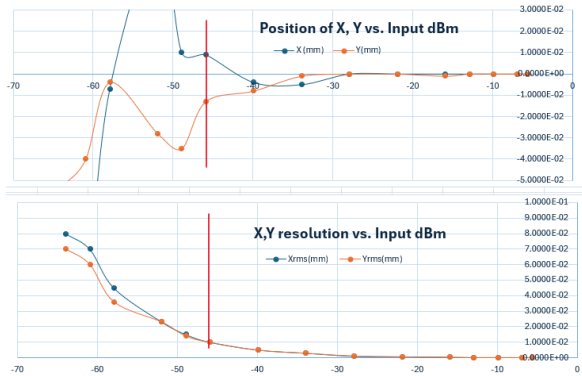

Fig. 6: Power scan: position, and resolution vs. input signal power

Channel crosstalk is only measurable in the IQ demodulated amplitude, which shows as high as 95 dB due to a selection of ultra-high isolation Mini-circuit RF switch HSWA2-63DR+. On the ADC raw data display, a typical noise level is less than 50 counts, which is 1.5 mV p-p since ADC input range is 2V p-p. The measured IQ demodulated amplitude shows a S/N of 75.6 dB is achieved by feeding a full range input signal and compared to an open channel.

*Position, Phase, and stability:*

With the BPM button signal, which is about 25% of the nominal beam signal, a position resolution less than 1 um (rms) is obtained, and a phase resolution of 0.1~ 0.2 (rms) degree is obtained by averaging the 1 us macro-pulse. See Fig. 7.

Temperature drift of the phase measurement had been addressed with special efforts. One is to have the RF clock signal to go through the exact same AFE channel as that of BPM signal, so both channels have the same temperature coefficient to keep the relative phase stable. This method was proven not enough due to the additional drift of the PLL LMK04832. Active self-calibrated method is introduced with a temperature sensor near the PLL chip and RF switch. With comparing the sensor reading against a pre-defined threshold $T_{sh}$, the firmware could automatically measure the phase of a copy of the RF reference clock to zero the drift. This active self-calibration could be performed during a no-beam period, since the beam duty cycle is only 1%. See Fig. 8, without calibration the measured phase drifts 2 degrees for 6 C degrees temperature change (10 C changes is the expected change of the electronics gallery annually), this drift is controlled to be within 0.2 degrees with the $T_{sh}$ set to 0.2 C.

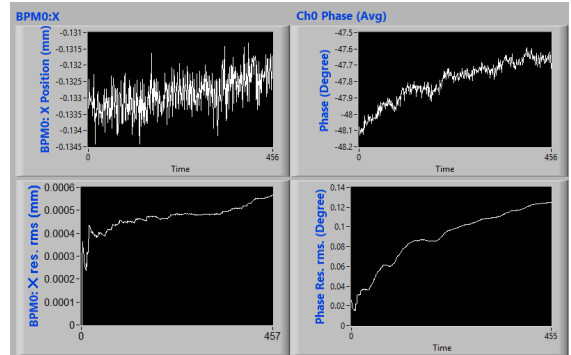

Fig. 7: Position and Phase Resolution for the BPM button signal

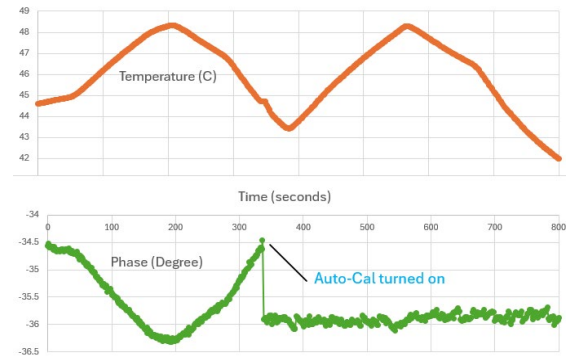

Fig. 8: Phase drift without and with auto-calibration

## CONCLUSION

A uTCA.4 based BPM electronics has been designed for the PIP-II BPM. A total of 9 pairs of AMC/RTM has been assembled to populate a 12 slot uTCA chassis to evaluate the full function of the system: aggregated 10 GbE data transfer, White Rabbit Timing, thermal performance, and uTCA management etc. The system performance: 10 um (rms) position resolution, 0.2 degree (rms) phase resolution, and 0.2 phase stability have been validated through a dedicated bench test system.

## REFERENCES

[1] PIP-II Conceptual Design Report, 2018,
   http://pip2-docdb.fnal.gov/cgi-bin/ShowDocument?docid=113;
[2] VadaTech AMC566 Hardware Reference Manual, November 19, 2023, Version 1.0,
   https://www.vadatech.com/product/amc566/;